# Extrinsic Anomalous Hall effect in Mn Doped GeSnTe Semiconductors in the Bad-Metal Hopping Regime

A. Khaliq,[1,*] R. Minikaev,[1] S. Zakar,[1] M. Arciszewska,[1] A. Avdonin,[1] V. E. Slynko,[2] and L. Kilanski,[1]

[1]*Institute of Physics, Polish Academy of Sciences, Aleja Lotnikow 32/46, PL-02668 Warsaw, Poland*
[2]*Institute of Materials Science Problems, Ukrainian Academy of Sciences, Chernovtsy, Ukraine*

We present high field magnetotransport studies of $Ge_{1-x-y}(Sn_xMn_y)Te$ bulk multiferroics with diamagnetic Sn and paramagnetic Mn concentration, $0.38 \leq x \leq 0.79$ and $0.2 \leq y \leq 0.086$, respectively. The zero-field resistivity, $\rho_{xx}(T)$ takes significant contribution from defects below $T \leq 20$ K however; a mixed scattering contribution from dynamic disorder and unusual sources is estimated from $T \approx 20$ K to 300 K. The carrier mobility shows anomalous temperature dependence from $\sim T^{-0.2}$ to $\sim T^{-0.5}$ which hints towards possible presence of polaronic effects resulting from coupling of holes with phonons. This anomalous behavior cannot be understood in terms of pure phononic scattering mechanism at high temperature. From point of view of high field results, the transverse component of magneto-resistivity manifests anomalous Hall effect originating from extrinsic scattering sources, particularly the side jump mechanism reveals a larger contribution. We also find that the correlation between the transverse and longitudinal conductivities follow the universal scaling law $\sigma_{xy} \sim \sigma_{xx}^n$ where $n = 1.6$ in the low conductivity limit. The values $n = 1.5 - 1.8$ obtained for the present GSMT alloys justify the bad-metal hopping regime since the results fall in the low conductivity ferromagnetic family with $\sigma_{xx} \leq 10^4$ $(\Omega$ cm$)^{-1}$. The interpretation of the $n = 1.6$ scaling in the low conductivity regime is thus far not fully understood. However; the anomalous Hall resistivity scaling with modified relation by Tian *et al.*, is indicative of the dominant side jump scattering along with noticeable role of skew scattering.

## Introduction

The ordinary Hall effect (OHE) discovered by Edwin Herbert Hall in 1879, revealed a transverse movement of charge carriers or electric field, $E_y$ caused by Lorentz force, $j_x \times B$ when exposed to a perpendicular magnetic field, $B$ [1,2]. Following the observation of charge carriers' deflection in a non-magnetic conductor, Hall recorded an unusual ten–fold larger effect when a ferromagnetic (FM) conductor was placed inside a perpendicular magnetic field [3]. Later named as extraordinary or anomalous Hall effect (AHE), was assumed to arise as a consequence of spin–orbit coupling (SOC) in FM phase having broken time reversal symmetry [4]. Unlike the OHE which varies linearly with the external magnetic field, AHE stems from spontaneous magnetization of the materials [5]. Pugh and Lippert formulated two contributions to $\rho_{xy}$ as; $\rho_{xy} = R_0 H_z + R_s M_z$, the first term is a contribution from OHE where $R_0$ is ordinary Hall coefficient whereas $R_S$ is anomalous Hall coefficient [6–8]. Since the discovery of AHE, its true origin has been discussed both by theoreticians and experimentalists for about a century due to non-existence of modern concepts e.g. topology and Berry phase mechanisms [4,9]. AHE has remained partially understood so far due to controversial and incomplete mechanisms spanning over a century. However; a successful set of concepts developed so far constitute of FM, SOC and disorder in the system [10].

In order to root out the origin of AHE, Karplus and Luttinger (KL) put forward a first theoretical approach who considered spin orbit interaction of polarized electrons in conduction band and interband mixing as sources of AHE [11]. The modern day Berry curvature which plays a crucial part in understanding AHE, was termed as "anomalous velocity" by Luttinger which yielded, $\rho_{int} \propto \rho_{xx}^2$ [3,11]. The KL theory was pioneered taking into account the intrinsic contribution to AHE rather than extrinsic such as scattering. Following KL theory, Smit proposed an extrinsic mechanism responsible for AHE which was centered on the asymmetric or skew scattering (SS) of the spin-polarized electrons induced by SOI [12,13]. The Smit's SS mechanism can be described by using classical Boltzmann equation as explained by E. A. Stern [14]. Another extrinsic contribution to AHE was proposed by L. Berger in 1970 who presented a side-jump (SJ) scattering of charge carriers off the impurities [15]. Here, the central potential displaces the center of mass of the wave packet sideways after scattering. In the context of extrinsic sources, $\rho_{xy}^{AH}$ varies linearly against longitudinal resistivity $\rho_{xx}$ when SS is dominant i.e. $\rho_{sk} \propto \rho_{xx}$ [16] however; a quadratic behavior is seen in case of SJ scattering, $\rho_{sj} \propto \rho_{xx}^2$ [16]. Apart from above scaling relations, several different approaches in terms of $\rho_{xx}$ were presented e.g. $\rho_{xy}^{AH} \propto b\rho_{xx}^2$ [17] $\rho_{xy}^{AH} \propto a\rho_{xx}$ in the low temperature range [18] $\rho_{xy}^{AH} \propto a\rho_{xx}+b\rho_{xx}^2$ [2] and $\rho_{xy}^{AH} \propto b\rho_{xx}^\alpha$ where $\alpha$ stays between 1 and 2 [19]. As previous works have recognized AHE to be proportional to net magnetization, Nakatsuji et al., claimed large room temperature AHE in antiferromagnetic $Mn_3Sn$ [20]. This suggests the existence of AHE is extraordinarily broader than previously thought and limited to ferromagnets only. As well, AHE finds applications in modern day spintronics such as development of memory devices [20].

In this article, we report the magnetotransport results and scaling of AHE results with a modified scaling law. The samples characterized in this work were grown by modified vertical Bridgman method. The present samples are a continuation of our previous work on $Ge_{1-x-y}(Sn_xMn_y)Te$ (GSMT) bulk multiferroic semiconductors [21−24]. The present alloys have large proportion of Sn varying between $0.38 \leq x \leq 0.79$ compared to the previous work whereas Mn is introduced in the range $0.02 \leq y \leq 0.086$ [21−24]. From crystal structure point of view, the sample with $x = 0.38$, $y = 0.02$ holds a rhombohedral (R3m) structure in coexistence with the cubic (Fm-3m) symmetry however; alloys with higher Sn+Mn contents switch to pure Fm-3m phase as discussed in Ref. 21,22. The Longitudinal resistivity, $\rho_{xx}(T)$ demonstrates contribution from static disorder at $T < 20$ K while phononic-like mechanism dominates at higher temperature. Moreover, the hole carrier mobility, $\mu_h(T)$ depicts strong temperature dependence in the lower temperature regime with peaks in $\mu_h(T)$ curves for Sn rich samples. The $\mu_h(T)$ results suggest that polar optical modes or polarons have probably dominant role in the high temperature regime. The scaling of AHE component is discussed in terms of $\rho_{xx}$ which identifies superposition of skew and side-jump scattering sources. Unlike the previous model defined in Ref. 2, the residual and dynamic disorder parts are separated which approximates the individual contributions using the modified scaling law presented in Ref. 16. Furthermore, the transverse magnetoresistance with negative magnitude is interpreted with the spin disordered model.

Energy-dispersive x-ray fluorescence (EDXRF) spectrometer provides an instantaneous technique to perform elemental composition analysis of alloys. The elemental proportions in GSMT bulk alloys were analyzed by EDXRF technique using Tracor x–ray Spectrace 5000 spectrometer. Afterward, the room temperature (RT) crystal structure analysis of GSMT alloys was executed using high resolution multipurpose XPert Pro MPD conventional x–ray diffractometer (HRXRD). The tube x–ray source used for these measurements works with $CuK_{\alpha 1}$ radiation having wavelength, $\lambda = 1.5406$ Å. In addition to structural analysis, the samples were subjected to magnetotransport measurements. Prior to the magnetotransport measurements, the alloys were cut into Hall bars of $1 \times 1 \times 8$ mm dimensions, six contacts of gold wire connected by using indium solder whereas a fixed current of 100 mA was applied during the whole measurement operation. In this part, both temperature and magnetic field dependent measurements were performed. The variable temperature measurements were made between $4.3 \leq T$ (K) $\leq 300$ with magnetotransport measurements at $|B| = \pm 1.4$ T. The high field longitudinal and Hall resistance results were obtained using a superconducting magnet up to $|B| = 13$ T. The variable field results were recorded down to $T \approx 1.6$ K and at selected temperatures depending on the magnetic phase transition of each composition.

The temperature dependent longitudinal resistivity, $\rho_{xx}(T)$ results obtained between $4.3 \leq T$ (K) $\leq 300$ are shown in Fig. 1(a,b). The $\rho_{xx}(T)$ graphs manifest a linear decrease from $T = 300$ K down to $\approx 20$ K which then remains almost independent of temperature down to $T \approx 4.3$ K. The monotonic temperature dependence of $\rho_{xx}(T)$ between $T = 20 - 300$ K represents a metallic-like behavior which likely occurs in degenerate semiconductors. At $T \geq 20$ K, $\rho_{xx}(T)$ presumably arises due to scattering of charge carriers by the dynamic disorder of the lattice. In the low temperature region, the residual resistivity, $\rho_{xx0(4.3\ K)}$ decreases tremendously from $\rho_{xx0(4.3\ K)} \approx 70$ mΩ cm for $x = 0.38$ to $\rho_{xx0(4.3\ K)} \approx 17$ mΩ cm for $x = 0.79$. The large $\rho_{xx0}$ values compared to other DMSs [25] indicate a substantial degree of atomic disorder in the alloys. The considerable variation in $\rho_{xx0}$ signifies that the low temperature resistivity originates due to scattering of charge carriers by the static disorder or impurity sites in the alloys. Such a large difference in electrical properties presumably depends on the doping content as reported before for $TiCo_xNi_{1-x}Sn$ [26]. Since the resistivity curves drop for high Mn content, this shows consistency with the previous results of inducing hole-carriers via doping [27]. The dissimilar number of induced defects also follows the deviation from Matthiessen's rule which illustrates that the slopes of all $\rho_{xx}$ vs $T$ curves should be parallel for the same type of alloys but with different content of doped ions. Such a deviation could be seen in terms of the diverse exponent values that denotes the curves slope, see the power law fitting in the following section. Further, the residual resistivity ratio (RRR), $\rho_{xx(300\ K)}/\rho_{xx(4.3\ K)}$ rises in heavily doped samples from 1.19 – 1.59 for $x = 0.38 – 0.79$, respectively, Fig. 1(b). The significantly large $\rho_{xx0}$ and extremely low RRR values are similar to previous GeTe based studies [28], however; quite large compared to other compounds e.g. $Co_2MnSi$, $NiMnSb$ Heusler alloys [29], $MgB_2$ single crystals [30] and $Sr_2Cr_3As_2O_2$ chromium oxypnictide layers [31]. The large $\rho_{xx0}(T)$ and low RRR of GSMT in this work are assumed to originate due to added impurities or atomic dislocations in the doped alloys [28]. This temperature independent part of electrical resistivity is generally a consequence of impurities, grain boundaries or dislocations which disrupt the periodicity of the lattice [32].

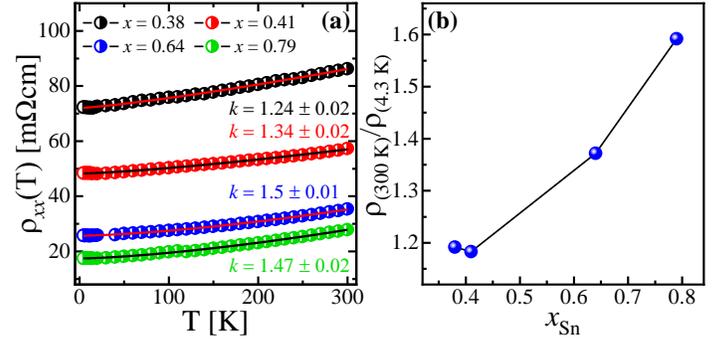

FIG. 1(a) Longitudinal resistivity, $\rho_{xx}(T)$ of the alloys with $0.38 \leq x \leq 0.79$ in the temperature range $4.3 \leq T \leq 300$ K. Solid lines denote fits to power law (eq. 1) in the entire temperature range. (b) Residual resistivity ratio, $\rho_{xx(300\ K)}/\rho_{xx(4.3\ K)}$ vs. Sn content.

In order to further understand the $\rho(T)$ behavior, we used power law given by eq. 1 to fit the experimental results.

$$\rho_{xx}(T) = \rho_{xx0} + CT^k \qquad (1)$$

As presented in Fig. 1(a), eq. 1 yields best fits in the entire temperature range shown as solid lines. For fitting procedure, $\rho_{xx0}$, $C$ and $k$ were set as free parameters where $C$ is coefficient of temperature and exponent $k$ denotes the slope of $\rho(T)$ curves. Here $\rho_{xx0}$ stems from the static impurities at low temperature while $CT^k$ comes from electron-electron collision or the phonon induced part that arises due to dynamic disorder. The residual resistivity, $\rho_{xx0}$ values were obtained very close to the experimental results with less than 3% deviation. The exponent, $k$ yielded value of $\approx 1.5(0.02)$ or $T^{1.5}$ dependence for highly doped samples whereas the low doped crystals display large deviation from $T^{1.5}$ and vary as $\rho_{xx} = f(T^{1.24 - 1.34})$. The $T^{1.24}$ dependence is to some extent close to pure phononic ($k = 1$) scattering with a slightly higher value. In the Sn-rich alloys, the $T^{1.5}$ dependence shows an extremely similar result obtained by Takagi et al., over a broad range of temperature [33]. In the context of the scattering sources, the true origin of such a temperature dependence of resistivity lacked adequate interpretation.

In Fig. 2, the isothermal curves of transverse magnetoresistance (MR), $\rho_{xx}(B)$ are presented which were obtained simultaneously with the Hall resistivity, $\rho_{xy}(B)$ results at $|B| = 13$ T. These curves were estimated after data were averaged for both the positive and negative currents taken at $I = 100$ mA. The final normalized results were determined using the relation, MR = $\Delta\rho_{xx}/\rho_{xx}(0) = (\rho_{xx}(B) - \rho_{xx}(0))/\rho_{xx}(0)$. Here $\rho_{xx}(B)$ and $\rho_{xx}(0)$ means resistivity values at $B = B$ and $B = 0$ field, respectively. All four alloys manifest negative MR at temperatures below $T_C$ point of each alloy.

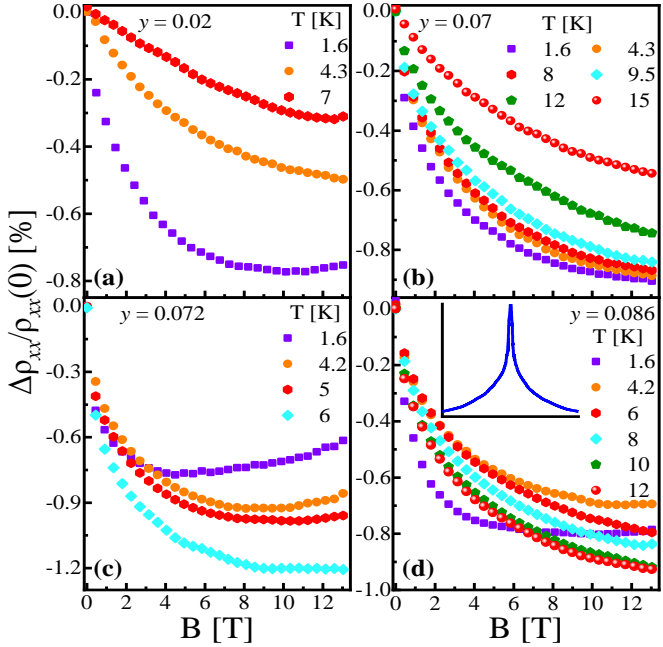

**FIG. 2** (a) $\Delta\rho_{xx}/\rho_{xx}(0)$ results at $T < T_C$ for each sample. The (a,b) and (c,d) exhibit two different trends as function of temperature where magnitude of MR decreases and increases, respectively. The inset of (d) shows a low-field part of MR at $T \approx 1.6$ K.

Depending upon the impurity content, these isotherms show two distinct classes, the low Sn/Mn doped samples in which the magnitude of MR decreases as temperatures rises and Sn/Mn rich alloys in Fig. 2(c,d) which reveal the opposite trend for MR(*T*) dependence. Except for the sample with *y* = 0.07, the MR curves at *T* ~ 1.6 K saturate for the remaining alloys at rather intermediate field of $|B| \approx 4 – 8$ T where (c) shows even a positive upturn in the slope, see Fig. 3(a) for comparison at $T \approx$ 1.6 K. This could also be interpreted as a broad minimum in MR around 5 T above which the MR magnitude commences to drop constantly up to 13 T. The maximum value of MR for each alloy is relatively small since the present GSMT alloys have low carrier mobility, (Carrier mobility results will be presented in Fig. 9) this might be the reason for such a small MR magnitude [34]. Besides the temperature dependence disparity, the magnitude of MR curves increases with the concentration of Mn impurity in the low field regime e.g. $|B| \leq 4$ T. The low field MR results for different Mn content are shown at *T* ~ 4.3 K, Fig 3(b). For the first three samples with *y* = 0.02, 0.07 and 0.072, the magnitude of MR seems to be an increasing function of impurity content. However; it drops for the highest impurity content of *y* = 0.086. This type of correlation between Mn impurity content and MR shows that the spin scattering process drops with increasing magnetic content which is in good agreement with the model and previous results [35]. The drop in MR of the sample with *y* = 0.086 however; could not be justified without detailed investigation of the two quantities. One possible explanation might be the charge scattering due to localization effect for this particular sample which produces a drastic variation near zero-field regime. As shown in the inset of Fig 2(d), this sample demonstrates MR with a similar behavior as a cusp-like shape which might indicate weak-localization effect at low temperature.

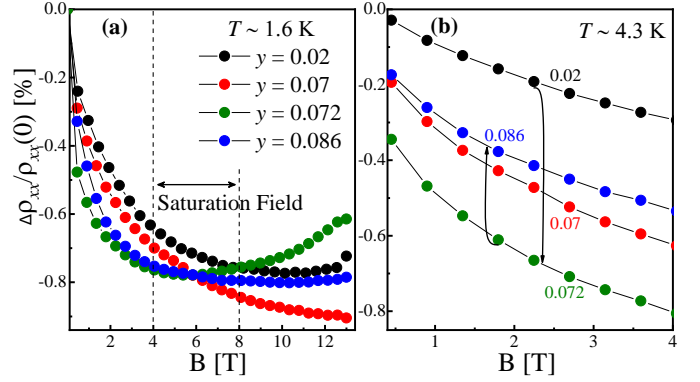

**FIG. 3**(a) $\Delta\rho_{xx}/\rho_{xx}(0)$ results at *T* ~ 1.6 K where dashed lines show approximate values of saturation field. (b) $\Delta\rho_{xx}/\rho_{xx}(0)$ results for the same samples shown at *T* ~ 4.3 K, the increase in MR as a function of Mn impurity level is presented except for *y* = 0.086 sample.

In addition, such a relation also indicates that the negative magnetoresistance stems from the presence of the magnetic impurity ions in the alloys [35]. However; one needs to be careful to accurately find out the origin of negative MR in the present alloys. A generally argued source of negative MR is the low temperature weak localization effect though it appears prominent in the weak field regime [36]. For the GSMT isotherms, the MR variation is significantly strong around zero-field regime which might be due the weak localization effect, particularly in Fig. 2(c,d). Also, the weak localization like behavior shown in Fig. 2(d) at low temperature is ascribed to the high degree of disorder in these alloys. The weak-localization effect is typically dominant around zero field and decays when the magnitude of magnetic field increases [23]. Additionally, negative MR is also attributed to the third order *s-d* exchange Hamiltonian during spin scattering process in disordered states [37,38] since the second order theory vanishes under strong magnetic field [38] in situations similar to Fig. 2 obtained at high field, $|B| = 13$ T. Furthermore, in the context of nature of scattering, the MR results are presented at $T \leq 15$ K which is within the residual resistivity limit of $T < 20$ K. This indicates that the role of phonons is negligible at temperatures at which MR results were obtained since $\rho_{xx0}$ is ascribed to static impurities. Also the $\rho_{xx0}$ contribution is presumed to be independent of applied magnetic field [39]. Consequently, the scattering of conducting charge carriers is assumed to occur exclusively from random magnetic impurities present in the GSMT system which leads to variation in MR as a function of the magnetic field.

The interpretation and scaling of negative MR fundamentally depends upon the orientation of magnetic moments in the alloy. The present samples have a broad range of impurity induced magnetic interactions as presented earlier [22]. These magnetometric results showed that GSMT alloys manifest frustrated magnetic states coexistent with small ferromagnetic-like clusters [22]. In similar diluted magnetic alloys such as CuMn, Monod had discussed that over a broad range of temperature and magnetic field, the amplitude of negative MR varied as a quadratic function of the magnetization [40]. However; deviations from the above quadratic behavior were also reported in case of concentrated alloys which were induced by the local magnetic field [40]. For Mn based canonical spin-glasses,

Majumdar previously presented a general scaling between $\Delta\rho_{xx}/\rho_{xx}(0)$ and magnetization [41],

$$\Delta\rho_{xx}/\rho_{xx}(0) = -\beta M^2 \qquad (2)$$

The variation of negative *MR* with the square of the magnetization yielded a universal curve in spin-glass systems. Here *M* denotes magnetization of the alloys and *β* is constant of proportionality. The variation of *β* with Mn impurities was shown to be unique as it has revealed both concentration dependent and independent behavior for different alloys [42,43]. In Fig. 4, the above mentioned correlation between $\Delta\rho_{xx}/\rho_{xx}(0)$ and magnetization is attempted for the current samples at $T \approx 4.3$ K and $|B| = 9$ T. The figure shows relatively fast variation of MR in the low field regime of less than $|B| \approx 3.5$ T whereas *M* exhibits large variation at high field. In Ref 44, the $\Delta\rho_{xx}/\rho_{xx}(0) \propto M$ showed a straight line with slope 2 over wide temperature range. In view of the above scaling relation, the current samples do not demonstrate such a universal scaling between MR and *M*. This might be indicative of a spin state different than conventional spin-glasses in which the magnetization $M = \langle S_i \rangle$ does not represent an order parameter. Here $S_i$ denotes magnetic moment associated with an $i^{th}$ impurity ion e.g. Mn in the GSMT matrix. Spin-glass systems manifest a disordered and frustrated spin-state below their transition temperature where both ferromagnetic and anti-ferromagnetic exchange interactions coexist. The absence of any obvious correlation between MR and *M* suggests a different interpretation for the current results. As this simplified scaling relation was formulated for canonical spin-glasses, therefore, such an interpretation holds inappropriate and deviates due to possible formation of small ferromagnetic clusters in the GSMT alloys.

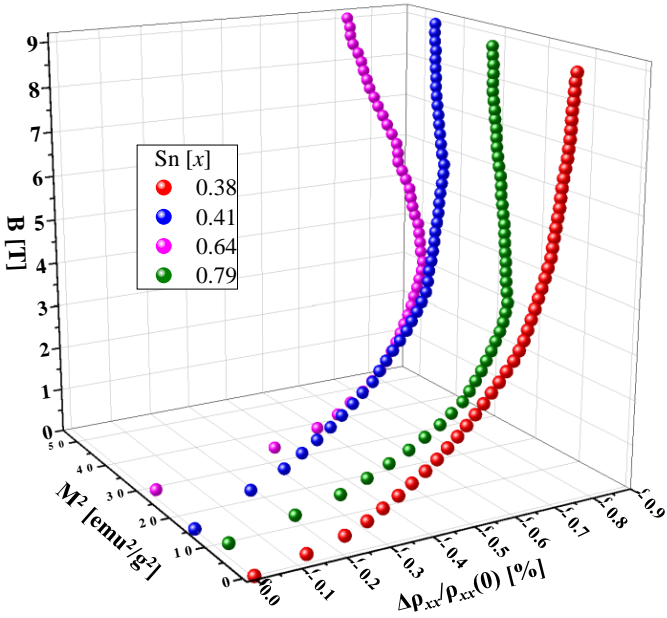

**FIG. 4** $\Delta\rho_{xx}/\rho_{xx}(0)$ results plotted against the square of magnetization following the relation $\rho_{xx} \propto -\alpha M^2$. The results are shown up to $|B| = 9$ T at $T \sim 4.3$ K.

The inception of negative *MR* in magnetic alloys could also be interpreted in the context of reduction of the spin dependent scattering as a consequence of spin alignment in the presence of magnetic field [45]. Thus, a spin-disorder model by Gennes and Fisher *et al.*, [46,47] is attempted to understand the negative MR in magnetically glassy GSMT alloys.

$$\rho_{sd} = 2\pi^2 \frac{k_F\, m^2\, \Gamma^2\, n_s}{n e^2 h^3} [S(S+1) - \langle S \rangle^2_{B,T}] \qquad (3)$$

Here $\rho_{sd}$ represents the resistivity term during the spin disorder scattering when magnetic field is applied. On the right side, the model takes contribution from electronic charge *e*, wave vector associated with Fermi level $k_F$, electron's mass *m*, Planck's constant *h*, density of electrons in the 3d state $n_S$, spin quantum number of paramagnetic Mn impurities is denoted by *S* and $\Gamma_S$ defines an effective factor extracted from the exchange integral associated with the charge carriers and magnetic impurities in the lattice. As discussed by Van Esch *et al.* the spin disorder scattering manifests a constant behavior in the paramagnetic region, however; it gradually drops when acting below $T_C$ until the complete alignment of the magnetic order [48]. The temperature independent behavior of spin disorder scattering in the paramagnetic region can be predicted by equation 3 since $\langle S \rangle = 0$ for $B = 0$. The fits to negative MR in Fig. 5 were obtained using eq. 4 in the spin-only ground state. Furthermore, a quadratic term, $dB^2$ was introduced in order to obtain the best fits where *B* is magnetic field and parameter, *d* is the coefficient of magnetic field.

$$\rho_{sd} = 2\pi^2 \frac{k_F\, m^2\, \Gamma^2\, n_s}{n e^2 h^3} \times$$
$$\left[ \frac{1}{2} + \left\{ \exp\left(\frac{-g_s \mu_B \mu_0 B}{2k_B T}\right) + \exp\left(\frac{g_s \mu_B \mu_0 B}{2k_B T}\right) \right\}^{-2} \right] + dB^2 \qquad (4)$$

Here $g_S$ is known as effective factor which is associated with the average value of the effective magnetic moment contributed by a Mn ion, $k_B$ is Boltzmann constant, $\mu_B$ denotes Bohr magneton, and *B* is applied magnetic field. The accurate interpretation of $\rho_{xx}(B)$ isotherms with the addition of quadratic term in eq. 4 suggests the presence of quadratic contribution to the *MR* results. This squared contribution possibly stems from the orbital motion of charge carriers in the presence of magnetic field [23]. In the present results, the coefficient of *B* was noted to vary with the Mn content during the fitting process. The quadratic term consistently contributed to the fitting process for all alloys whereas excluding this component did not reproduce the experimental results accurately. Finally, the free parameter $g_S$ (magnetic moment) values were extracted for all isotherms in Fig. 5(a–d). The obtained values varied between $g_S \approx 2.1(\pm 0.4)$ to $8.1(\pm 0.6)$ at $T \approx 1.6$ K. These values do not show a proper correlation with the Mn impurity concentration e.g. $g_S \approx 4.1$ for $y = 0.086$ whereas $g_S \approx 8.1$ was estimated for $y = 0.072$ indicates a drop in the magnitude of magnetic moment for higher Mn level. The fitting parameter values obtained for all samples are provided in table I. As discussed previously for the same alloys [22], the magnetic interactions in these glassy alloys dramatically vary depending both on the diamagnetic Sn and paramagnetic Mn ions in the lattice. The unlikely decrease of magnetic moment with higher Mn ions was assumed to partially arise due to Sn content. It is possible that the Sn-concentrated alloys influence the lattice constants which consequently tune the interaction between Mn ions in an unequal pattern. Apart from this, the randomly distributed Mn ions might also result into significant variation of the magnetic moment. Furthermore, the mixed nature of magnetic interactions in these alloys due to spin disordered or cluster state could have robust impact on the magnitude of magnetic moment as well. In such diluted magnetic alloys, it appears that not all Mn ions participate equally in inducing the magnetic order in the system.

**TABLE I:** Free parameter, $g_S$ values determined by fitting the MR curves with eq. 4. The table shows values obtained at $T = 1.6$ only.

| Sn | Mn | $g_S \pm \Delta g_S$ [1.6 K] |
|---|---|---|
| 0.38 | 0.02 | 2.1±0.4 |
| 0.41 | 0.07 | 2.6±0.4 |
| 0.64 | 0.086 | 4.1±0.4 |
| 0.79 | 0.072 | 8.1±0.4 |

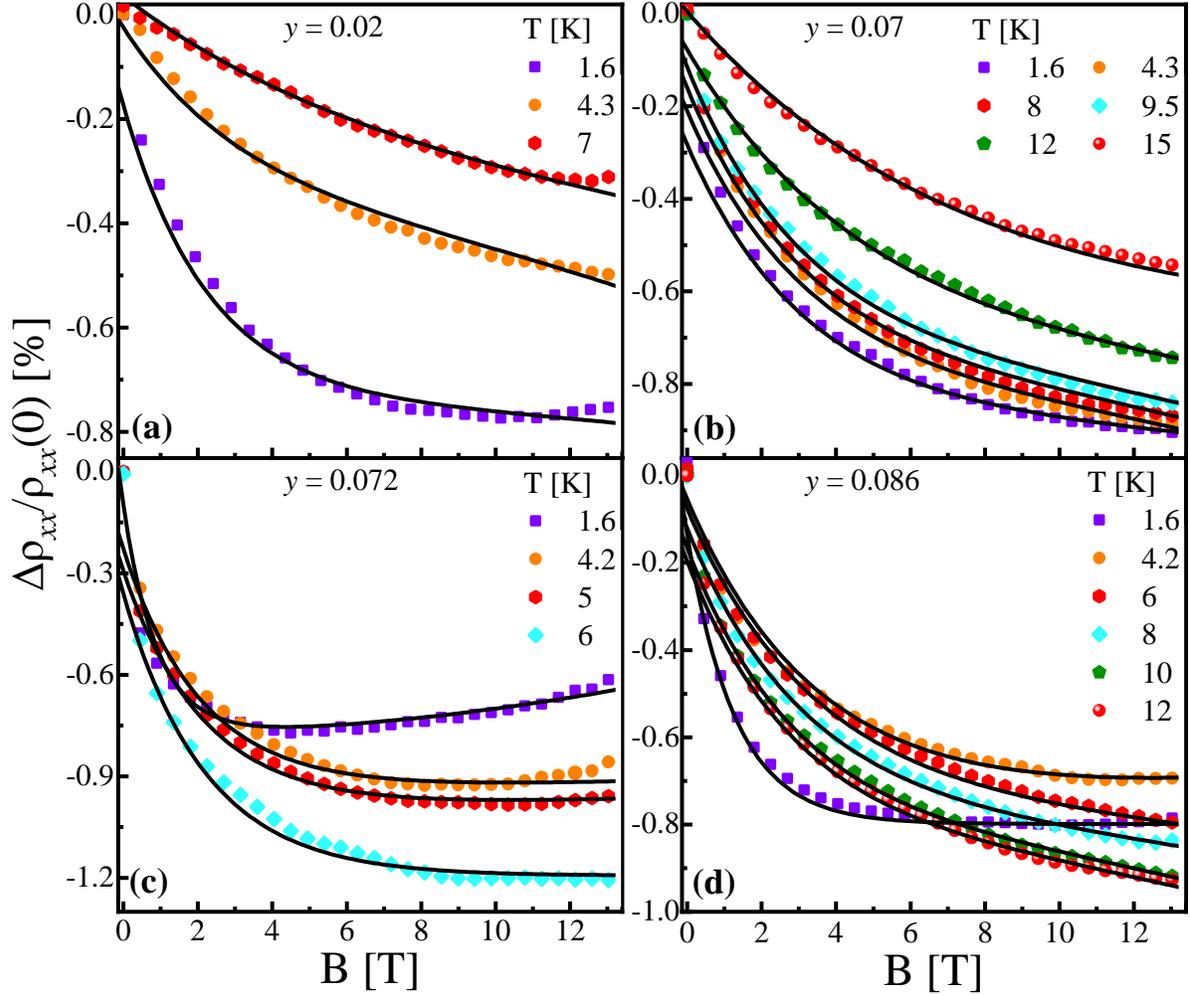

**FIG. 5** Isotherms of the negative magnetoresistance, *MR* measured below the magnetic transition temperatures. The experimental data shown as scatter-points is fitted (solid lines) to the spin-disordered model given by eq. 4.

In this section, the Hall resistivity component, $\rho_{xy}$ of the magnetotransport results measured at $|B| = 13$ T is discussed. Several isothermal $\rho_{xy}(B)$ results are presented in Fig. 6 in order of increasing Mn content from $y \approx 0.02 - 0.086$. All alloys demonstrate positive $\rho_{xy}$ isotherms up to $|B| = 13$ T with decreasing magnitude against increasing temperature. For the lowest Mn doping of $y \approx 0.02$, a small AHE effect could only be seen at $T \approx 1.6$ K due to its spin-disordered state at $T \approx 4.5$ K [22] and above. In comparison, the alloys with comparatively higher $T_C$ manifest large AHE effect, Fig. 6(b–d). The large magnitude of anomalous $\rho_{xy}$ in the low field region is limited to $|B| = 0.2 - 0.3$ T which then develops into ordinary linear dependence up to 13 T. The two components of $\rho_{xy}(B)$ are generally interpreted by the equation, $\rho_{xy} = R_0 H_z + R_s M_z$ as the field dependent OHE and magnetization dependent AHE parts, respectively. As interpreted by Smith and Sears [49], The OHE and AHE components in Fig. 6 seem to be consistent with the magnetization, $M(B)$ curves for the same samples, see Ref. 22. Obviously the $\rho_{xy}(B)$ and $M(B)$ curves demonstrate excellent correspondence which means that AHE could be well interpreted by the relation $\rho_{xy} = R_0 H_z + R_s M_z$.

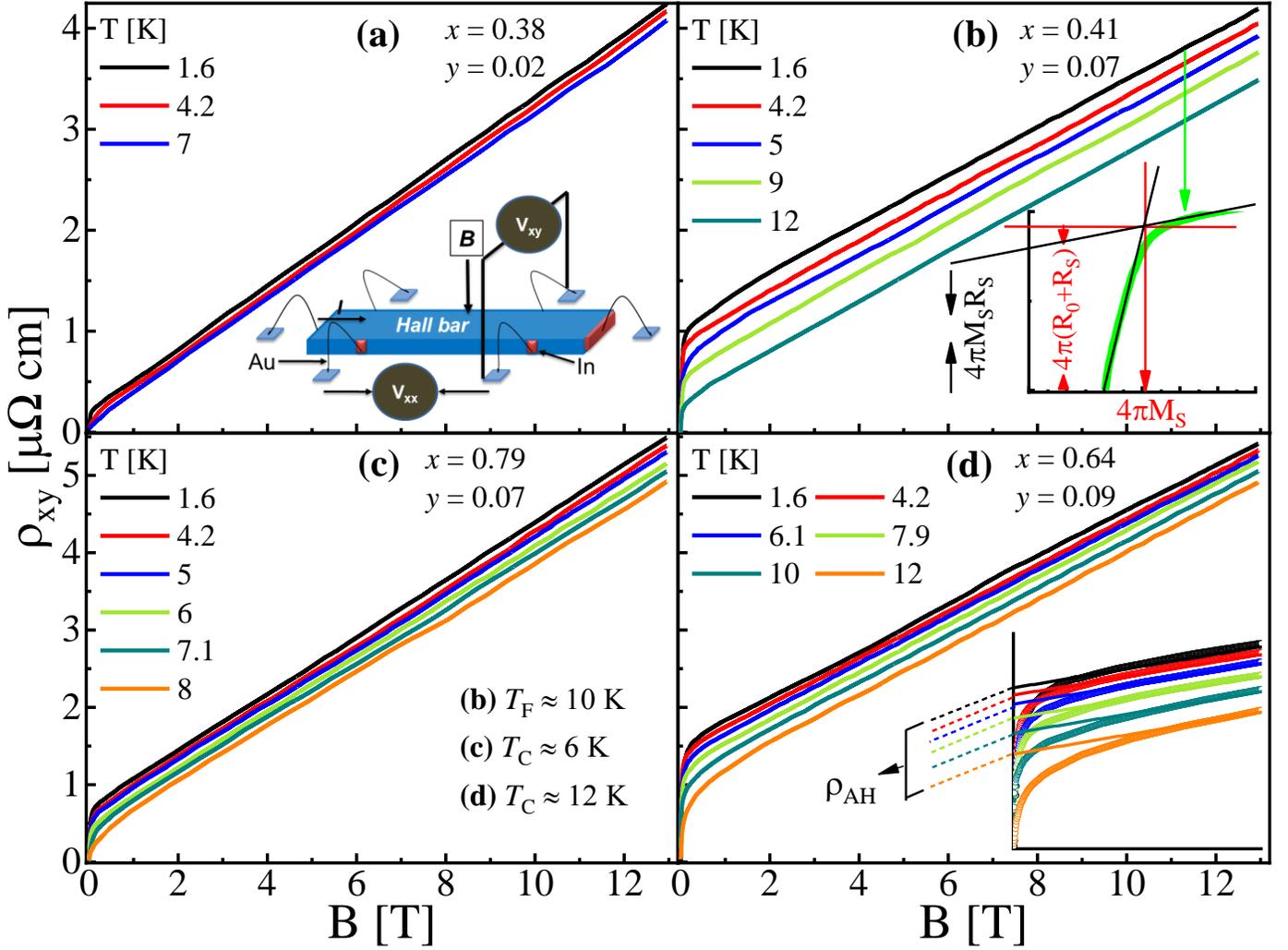

**FIG. 6**(a – d) Temperature dependent Hall resistivity, $\rho_{xy}(B)$ results for GSMT alloys with different Sn and Mn contents. The inset of (a) is a schematic presentation of the Hall bar connected to the holder via six contacts dc method. Inset of (b) depicts the estimation of OHE, AHE components and $M_S$. In (c), the freezing temperature, $T_F$ of magnetically glassy alloy and ferromagnetic-paramagnetic temperature, $T_C$ are provided. The inset of (d) shows linear fits of high field region which yield anomalous Hall resistivity, $\rho_{xy}^{AH}$ for each isothermal curve. The anomalous behavior of $\rho_{xy}(B)$ is consistent with magnetization and $T_C$ points presented in Ref. 22.

In the inset of Fig. 6(b), a low field cut of the $\rho_{xy}(B)$ isotherm at $T \approx 1.6$ K is presented; this shows an anticipated linear part as a function of $B$ and a weakly dependent part of $\rho_{xy}(B)$ with small gradient. The black solid lines indicate extrapolated lines to AHE and OHE parts of $\rho_{xy}(B)$ where $y$-intercept estimates spontaneous Hall effect denoted by spontaneous Hall coefficient, $R_S$ that is characteristic of magnetic materials and spontaneous magnetization, $M$. The solid red lines drawn through the extrapolated lines determine saturation magnetization, $4\pi M_S$ and total Hall resistivity, $4\pi(R_0+R_S)$ at $x$ and $y$-intercepts, respectively [50].

In order to identify the dominant scattering sources that cause the AHE part, it is fundamentally essential to correlate the transverse Hall resistivity to the longitudinal resistivity [4,16,51]. For the GSMT alloys, the scaling analysis of $\rho_{AH}$ in terms of $\rho_{xx}$ is presented in Fig. 7 to parse the major source of AHE. The first approach to find out the origin of AHE results is discussed via correlation between $\rho_{AH}$ and longitudinal resistivity, $\rho_{xx}$. Prior to the scaling relations shown in Fig. 7, basic linear and quadratic correlations, $\rho_{AH} \propto \rho_{xx}$ and $\rho_{AH} \propto \rho_{xx}^2$ were attempted however; convincing fittings were not achieved. This probably means that the origin of AHE in the present alloys is rather complex that could not be deduced from a single scattering mechanism. The above two correlations are generally used to distinguish between the SS and SJ scatterings which cause AHE. In order to identify conclusive interpretation, the scaling analysis was extended to relations that take into account both the SS and SJ mechanisms. In Fig. 7(a), $\rho_{AH}$ as a function of $\rho_{xx}$ was fitted to a simplified relation, $b\rho_{xx}^\alpha$ where the exponent $\alpha = 1$ denotes SS and $\alpha = 2$ means contribution from scattering dependent SJ and scattering independent intrinsic process induced by SOI [11,15]. This simple model could either separate SS or SJ or identifies superposition of scattering processes depending on the outcome of parameter $\alpha$. The free parameter values from Fig. 7(a) as presented in table II were obtained as $|\alpha| = 1.75$, 1.8 and 1.6, for $y = 0.07$, 0.072 and 0.086, respectively. Since $\alpha$ yields values from 1 – 2 for these results, it suggests an intermediate range where contributions are assumed to come from superposition of SS and SJ scattering processes [52]. These intermediate values might not sort out the true contribution proportion of each mechanism however; it could be interpreted that the SJ scattering is dominant here as deviation of α is large from the SS scattering (α = 1). It is worth mentioning that these values do not always hold the narrow range of α = 1 − 2 and could exceed beyond 2. In the regime α > 2, the non-validity of this scattering relation was argued to come from larger mean free path compared to layers' thickness [53]. In another example, $\rho_{xx}^{3.9}$ was obtained for granular alloys where α = 3.9 resulted due to scattering rate at the interface [54] or due to residual resistivity, $\rho_{xx0}$ in γ′-Fe$_4$N films where the SJ and intrinsic mechanisms are dominant [55]. These discrepancies in α > 2 regime suggest the exponent α does not hold universal values but fluctuates with geometry of the system e.g. thickness or granule's size. Besides the distinction of scattering mechanism on the basis of α, in general, asymmetric SS shows dominance at low temperature whereas SJ takes over in the high temperature range [39]. The AHE behavior exhibited by the current samples exists in the low temperature regime which on the basis of the above initial scaling does not originate from a unique and clearly dominant scattering mechanism. This outcome will be further justified using a comparatively new scaling law in the following section.

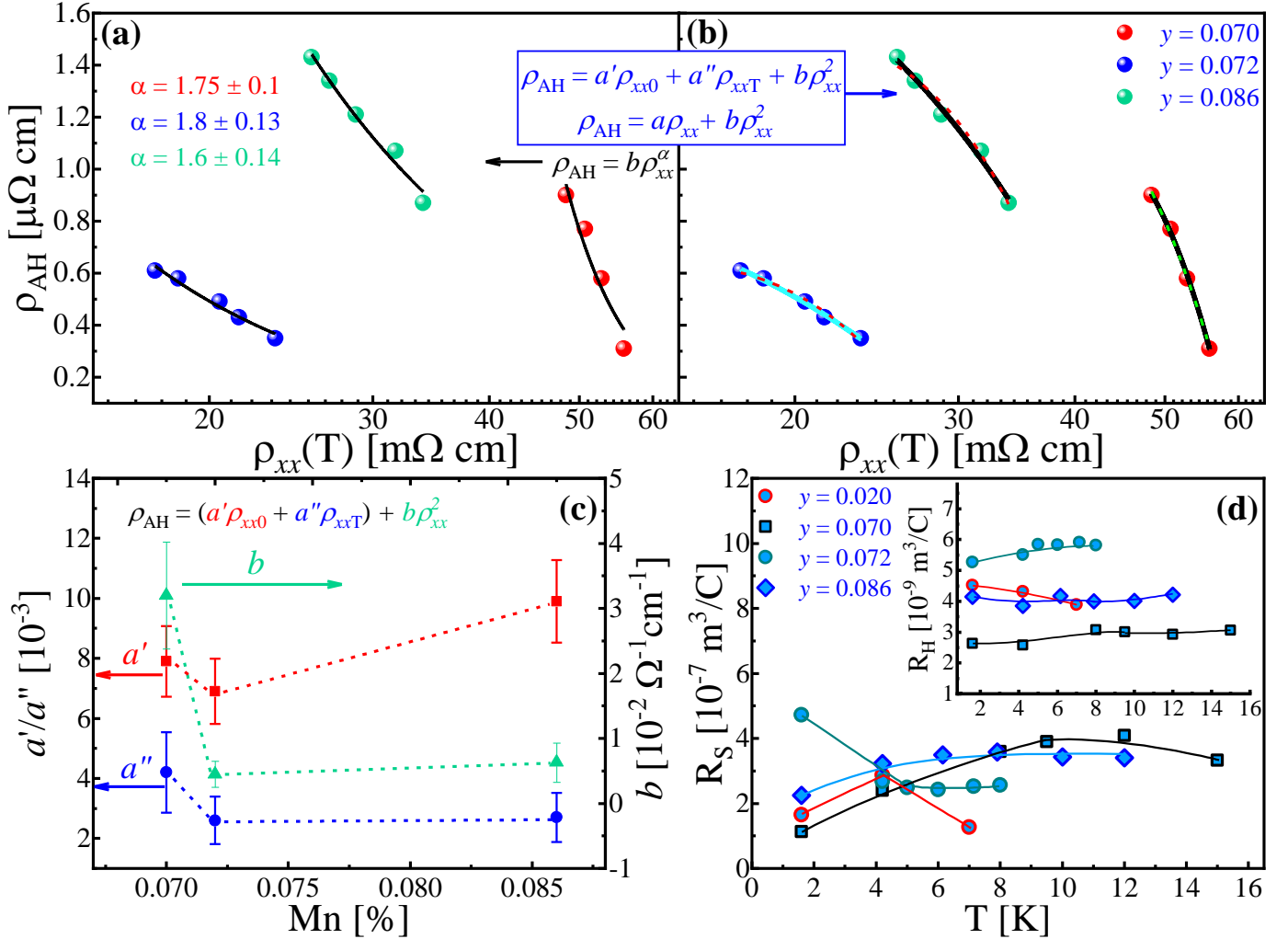

**FIG. 7** (a) Scaling of anomalous Hall resistivity, $\rho_{AH}$ vs longitudinal resistivity, $\rho_{xx}$ with the simplified relation indicated by black arrow. (b) Fitting results of $\rho_{AH} = f(\rho_{xx})$ with additive two-term relation and modified scaling law marked blue. Comparison of previous and modified relations is shown as dashed and solid lines, respectively. (c) Obtained parameters of the residual resistivity $a'$, phonon induced resistivity, $a''$ and side jump or intrinsic mechanism $b$ as a function of Mn content. (d) Anomalous Hall constant, $R_S$ obtained from isothermal results of figure 6. The inset presents ordinary Hall coefficient, $R_0(T)$. Solid lines in (d) are drawn as guide to the readers.

**TABLE II:** Fitting parameters $a'$, $a''$, $b$ and α extracted after scaling the results with the relations $\rho_{AH} = f(\rho_{xx})$ and $\rho_{xy}^{AH} \propto b\rho_{xx}^{\alpha}$, respectively.

| Mn | $a'(10^{-3})$ | $\Delta a'$ | $a''(10^{-3})$ | $\Delta a''$ | $b(10^{-2})$ | $\Delta b$ | α | Δα |
|---|---|---|---|---|---|---|---|---|
| 0.07 | 7.9 | ±1.18 | 4.2 | ±1.3 | 3.2 | ±0.82 | 1.75 | ±0.1 |
| 0.072 | 6.9 | ±1.09 | 2.6 | ±0.79 | 0.45 | ±0.2 | 1.8 | ±0.13 |
| 0.086 | 9.9 | ±1.4 | 2.7 | ±0.82 | 0.63 | ±0.31 | 1.6 | ±0.14 |

In Fig. 7(b), a two component additive relation of SS and SJ terms, $\rho_{AH} = a\rho_{xx} + b\rho_{xx}^2$ is attempted to further evaluate whether the above interpretation holds accurate. The SS parameter is denoted by $a$, whereas $b$ contains contributions from both SJ and intrinsic mechanism. The fits in Fig. 7(b) show that the experimental results could be properly described by the additive relation of both the scattering mechanisms. However; this equation does not uniquely separate the SS and SJ components, e.g. Kondorskii et al. and Crepieux et al. have shown that the SS component also contributes to the quadratic part even when properly described [56,57]. The above generalized equation was recently modified by Tian et al., who proposed the SS term as a sum of the residual resistivity, $a'\rho_{xx0}$ which is defects dependent caused by static impurities and the temperature dependent part, $a''\rho_{xxT}$ ($\rho_{xxT} = \rho_{xx} - \rho_{xx0}$) that is proportional to dynamic disorder related scattering induced by phonons or magnons [16,51]. The splitting of the total resistivity into residual and phonon induced parts originally comes from the Matthiessen's rule, $\rho_{xx} = \sum_i (\rho_i)$ in which $\rho_i$ represents the resistivity component by the $i$th kind of disorder scattering [58,59]. Fits shown as solid lines in Fig. 7(b) were obtained with the modified scaling law, $\rho_{AH} = a'\rho_{xx0} + a''\rho_{xxT} + b\rho_{xx}^2$ which treats the SS part as two independent scattering sources. Besides the outcome of different scattering parameters, the fitting results of the modified scaling law (solid lines) are compared to those obtained with the previous generalized model (dashed lines). The modified model clearly yields superior fits compared to both the previous models in Fig. 7(a,b). The free parameters $a'$ and $a''$ could be estimated to make a distinction between the defect dependent and dynamic disorder induced components of $\rho_{xx}$ whereas the magnitude of $b$ interprets the role of SJ or intrinsic mechanism [15]. Moreover, the separation of intrinsic and SJ mechanisms is challenging as both of them correlate in a quadratic way to $\rho_{xx}$. Since the role of phonon scattering is generally very small in the case of SS mechanism, the modified scaling law is also aimed at separating the roles of residual and temperature dependent resistivity, $\rho_{xx0}$ and $\rho_{xxT}$, respectively. However; the key outcome would be to compare the parameter $b$ with the earlier scaling results Fig. 7(a) and hence validate whether the contribution of SJ scattering is still dominant. The main difference in the current work is that the variation of free parameters is examined as a function of impurity concentration rather than thickness dependent results as discussed in Ref. 16.

The obtained results of free parameters are shown in Fig. 7(c) for three alloys in (b); the extracted values and uncertainties are presented in Table II. The results of $a''$ and $b$ extract large values for low impurity content whereas $a'$ does not show any obvious trend. First important outcome of the scaling reveals the parameter, $a''$ of $\rho_{xxT}$ is smaller than that of $\rho_{xx0}$ component. This agrees with the general notion that the phonon induced contribution is indeed smaller than the residual part of SS mechanism [16]. However; the non-negligible values of $a''$ compared to $a'$ means the overall role of SS part comes from mixed sources of both phonon-induced and residual resistivity. In cases where phonon scattering role is trivial, the $\rho_{xxT} \simeq 0$ approximation could be made in the zero temperature limits in which the total resistivity becomes $\rho_{xx} \simeq \rho_{xx0}$ [58]. The second central result is the comparison of SS and SJ contributions to $\rho_{AH}$ obtained from the new scaling model. Here the SJ parameter, $b$ shows values an order of magnitude larger than the SS parameter, $a'$ that satisfies the earlier discussion of dominant SJ contribution. On the other hand, the SS parameter reveals smaller but non-zero values which could be interpreted as the SJ source is dominant however; role of SS cannot be ruled out. Based on the fitting results of GSMT alloys, the interpretation of $\rho_{AH}$ with the new scaling law and the one in Fig. 7(a) both find that extrinsic sources are dominant which take large part from SJ and extract minor contribution from SS. In case of the SJ scattering, as described by L. Berger [15], the wavefunction of the charge carrier suffers a lateral displacement, $\Delta l$ with respect to the scattering center. The scattering centers in the current samples are Mn impurities which act as central potential sites. The value of $\Delta l$ was estimated to be around 1Å according to the relation $\Delta l = -\lambda_{S.O} \sigma_z k_x$. Here $\lambda_{S.O}$ denotes the spin orbit coupling parameter; $\sigma_z$ is the spin value of the charge carrier whereas the $x$–component associated with Fermi wave vector is denoted by $k_x$. The notion of $\Delta l$ holds particular significance in ferromagnetic alloys similar to the GSMT materials and metals since the mean free path of the charge carriers is of the order of $\Delta l = 1$Å after being scattered by magnetic impurities [15]. Due to the relatively low mobility of the GSMT DMSs, the mean free path is assumed to have comparable range as the average separation between Mn ions in the system [60].

In Fig. 7(d), the temperature dependence of anomalous and ordinary Hall coefficient, $R_S$ and $R_H$ are shown obtained as free parameters after fitting $M(B)$ to $\rho_{xy}(B)$ isotherms. Here, $R_H$ arises from the Lorentz force while on the contrary $R_S$ stems from spontaneous magnetization e.g. ferromagnets below $T_C$. Above $T_C$, the spin disordered state produces insignificant Hall current however; it increases significantly below $T_C$ due to alignment of spins. Such a large Hall current arises due to left-right asymmetric scattering induced by spin-orbit coupling between the charge carriers and the lattice [61]. Apparently, AHE contribution is dominant over OHE since $R_S(T)$ is two orders of magnitude larger than $R_H(T)$ at $T \leq 15$ K. This magnitude of $R_S(T)/R_H(T)$ is consistent with our previous work on GSMT and Ge$_{1-x-y}$Mn$_x$Eu$_y$Te bulk alloys [23,62]. The positive sign of both $R_S(T)$ and $R_H(T)$ indicates the transport mechanism is controlled by holes as majority charge carriers. An interesting outcome here is the nearly temperature independent behavior of $R_S/R_H$ below their magnetic transition temperatures. This profile is similar to results of other IV-VI narrow gap SCs e.g. Mn and rare earth ions incorporated into SnTe alloys [63]. However, the temperature dependence and transition to negative regime of $R_S$ was earlier recorded for the same alloy of Sn$_{1-x-y}$Eu$_x$Mn$_y$Te [64] which later disappeared under high field [62]. In addition to IV-VI SCs, the weakly $T$-dependent nature of $R_S/R_H$ was also obtained in GaMnAs, selected samples of Fe and Mn based silicides and Fe$_3$GeTe$_2$, a ferromagnetic van der Waals semimetal [65,66]. The negligible influence of temperature on $R_S$ signifies the possible potential of GSMT alloys for developing Hall devices. However; $R_S(T)$ seems to have stronger temperature dependence below $T \approx 4$ K and drops faster than $R_H(T)$, which stays almost constant even at lowest temperature. It is also seen that the variation of $R_H/R_S$ with concentration of dopants ions of Sn/Mn is not demonstrating any obvious dependence. This might happen due to the fact that different content of diamagnetic Sn probably influences the magnetic interactions among Mn ions. Such an explanation could be justified based on the tuning of magnetic interaction by Sn reported for the same alloys [22]. However, the true mechanism behind the complex dependence of $R_H/R_S$ might arise from other parameters and could not be only attributed to impurities.

In the following section, the scaling relation between longitudinal and anomalous Hall conductivity (AHC) components $\sigma_{xx}$, and $\sigma_{AH}$ is discussed. Both the quantities were estimated using $\sigma_{xx} = \rho_{xx}/(\rho_{xx}^2+\rho_{AH}^2)$ and $\sigma_{AH} = \rho_{AH}/(\rho_{xx}^2+\rho_{AH}^2)$ tensor equations which might be reduced to $\sigma_{xx} = \rho_{xx}/\rho_{xx}^2$ and $\sigma_{AH} = \rho_{AH}/\rho_{xx}^2$ when $\rho_{xx} \gg \rho_{AH}$ [67]. For the conductivity scaling, a comprehensive theory was recently proposed by Onoda et al., that assesses both intrinsic and extrinsic AHE in multiband doped ferromagnets [68]. Here the conductivity scaling is generally categorized into three broad regions, (a) superclean metal regime $\sigma_{xx} > 10^6$ $\Omega^{-1}$ cm$^{-1}$ where linear dependence diverge as $\sigma_{AH} \sim \sigma_{xx}$ represents superior SS source (b) moderately dirty regime $\sigma_{xx} > 10^4 - 10^6$ $\Omega^{-1}$ cm$^{-1}$ which is scattering independent dominated by intrinsic mechanism and (c) dirty metal regime $\sigma_{xx} < 10^4$ $\Omega^{-1}$ cm$^{-1}$ in which the scaling relation follows $\sigma_{AH} \sim \sigma_{xx}^{1.6}$. The scaling relation between $\rho_{AH} \sim \rho_{xx}(\rho_{xx}^2)$ is assumed to yield nearly similar outcome as $\sigma_{AH} \sim \sigma_{xx}(\sigma_{xx}^2)$ [69]. The obtained results of the $\sigma_{AH} = f(\sigma_{xx})$ scaling are presented in Fig. 8 at the same temperatures as in Fig. 6. It is obvious that the $\sigma_{xx}$ results are within the bad-metal–hopping regime ($\sigma_{xx} < 10^4$ $\Omega^{-1}$) which allows the interpretation of the form $\sigma_{xy}$ = const. $\sigma_{xx}^n$ [4,70].

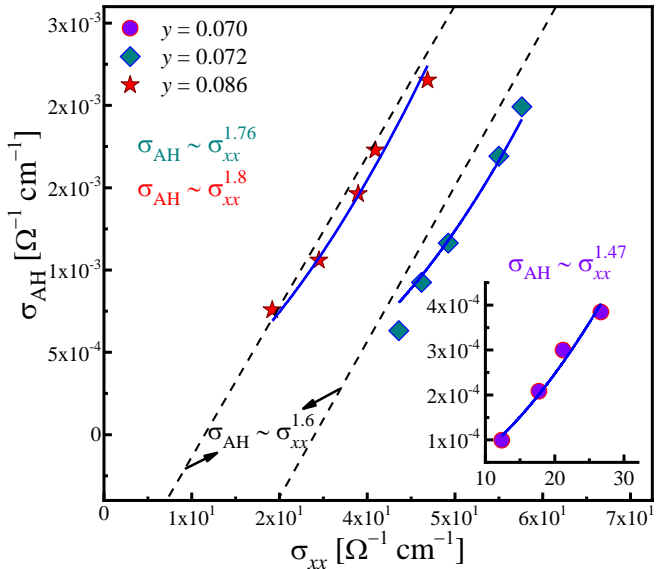

**FIG. 8** Scaling between magnitude of anomalous Hall conductivity, $\sigma_{AH}$ and longitudinal conductivity, $\sigma_{xx}$. The scaling relation, $\sigma_{AH} \sim \sigma_{xx}^n$ yielded fits shown as solid-blue lines whereas dashed-black lines denote reference fits for two alloys with $n = 1.6$.

The scaling in Fig. 8 to $\sigma_{xy} \sim \sigma_{xx}^n$ shows good fits shown by solid blue lines over the entire range of $\sigma_{xy} = f(\sigma_{xx})$ results though deviation from universal line $n \sim 1.6$ could be seen. The estimated values, $n = 1.76$ and $1.8$ are within the previously established limit of universal scaling, $n = 1.33 - 1.8$ for majority of experimental results [4,71] whereas $n = 1.47$ is close to the lower boundary. This non-trivial exponent, $n \approx 1.6$ has been experimentally achieved in the poorly conducting regime for many systems such as itinerant ferromagnets [67], nanogranular films [72], ferromagnetic van der Waals semimetal [66], $\varepsilon$-Fe$_3$N nanocrystalline films [73] and several other different systems [67]. The fact that $\sigma_{xy} \sim \sigma_{xx}^n$ scaling did not follow a linear relation ($n = 1$) or $\sigma_{xy} \sim \sigma_{xx}^0$ where $\sigma_{xy}$ holds constant value further supports the low conductivity hopping regime in GSMT alloys that is consistent with theoretically predicted value of $n = 1.6$. Previously, the parsing of AHE sources in these regimes has been discussed mainly into intrinsic and extrinsic mechanisms. As the crossover occurs from linear to constant regime of $\sigma_{xy}$, the SS source decays whereas intrinsic and SJ act as major AHE sources [68,44]. Typically, the intrinsic deflection mechanism is induced in systems when the multiband SOI system is linked to momentum space Berry phase [71]. But the contribution of SJ is much smaller than the intrinsic source in the plateau region which could be ignored as argued theoretically by Miyasato el al., [44]. Since $\sigma_{xy}$ = constant and $n = 1$ regimes cannot be seen in Fig 8, it is safe to disregard the dominant role of both Berry curvature and asymmetric SS sources in the present alloys. Here in the hopping regime, the intrinsic part initiates a damped decay due to impurities which follow a universal $\sim\sigma_{xx}^{1.6}$ dependence [67]. Furthermore, $\sigma_{AH}$ is significantly suppressed in dirty regime compared to $\sigma_{xx}$ due to damping or resonating condition [66]. This universal scaling dependence in hopping regime is remarkable as it yields same correlation irrespective of temperature, pressure and impurity content though depends on magnitude of $\sigma_{xx}$ and impurity potential [68]. It appears that the phonon scattering and static disorders do not substantially influence the charge-carrier dynamics and therefore yields universal scaling [69]. Thus, the control of $\sim\sigma_{xx}^{1.6}$ scaling reveals that spin-polarized charge carriers play vital role in inducing ferromagnetism that lead to AHE in these alloys. Also, in terms of nature of hopping, this scaling maintains its universality within $n = 1.33 - 1.76$ limits whether the system is in Mott variable-range hopping, interactions dependent or nearest-neighbor hopping [71].

The hole-carrier mobility, $\mu_h$ and carrier concentration, $n_h$ as a function of temperature are presented in Fig. 9. Drastic decrease in $\mu_h(T)$ curves shows significant temperature dependence below $T \approx 50$ K while exhibit extremely small variation between $T \approx 50 - 300$ K. The two samples with high impurity contents; $x = 0.64, 0.79$ and $y = 0.086, 0.072$ show increase in $\mu_h(T)$ results from $T \approx 4.3$ K up to $T_{max} \approx 13$ K and 9 K, respectively, at which maxima were recorded. Above the maxima, both samples manifest decline in $\mu_h(T)$ up to $T \approx 300$ K. The temperatures at which maxima occur for the two crystals are marked as $T_{max}$, see Fig. 9(a,b). On the contrary, the samples with low impurity contents do not show any peaks. The maxima in $\mu_h(T)$ curves separate two different charge scattering regimes. At temperatures below $T_{max}$, scattering from ionized impurities have dominant contribution whereas temperature dependent lattice scattering dominate at $T > T_{max}$ [74,75]. The impurity induced scattering at $T < 15$ K is in good agreement with the $\rho_{xx}(T)$ results. For the lattice induced part in the range $T > T_{max}$, detailed analysis will be presented in the following paragraphs.

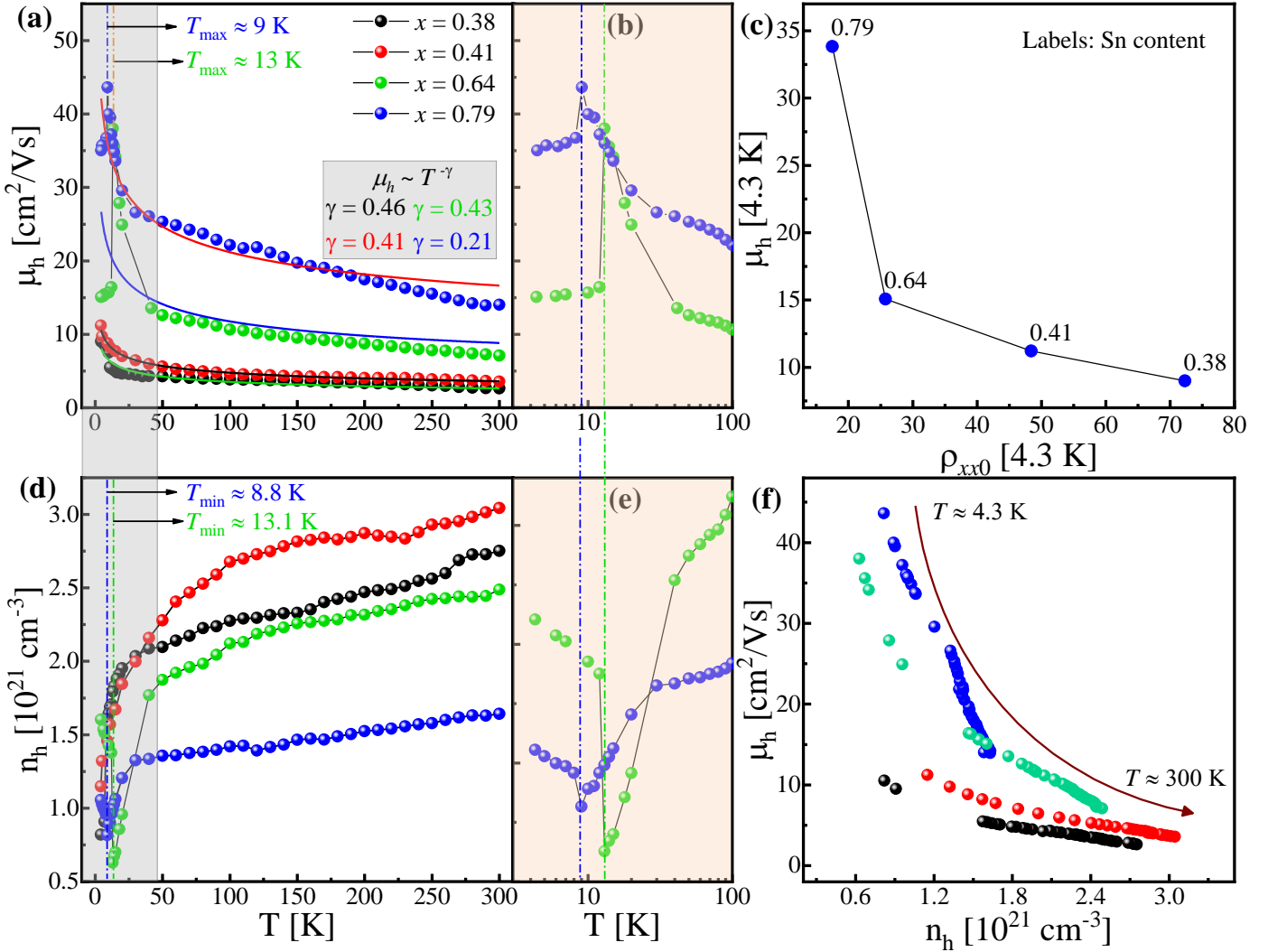

**FIG. 9** (a) The mobility, $\mu_h(T)$ results between $T \approx 4.3$ K – 300 K where solid lines denote fits to $\mu_h \sim T^{-\gamma}$. (b) Low temperature magnified part of (a) in order to visualize the $\mu_h(T)$ maxima for the samples with $x = 0.64$ and $0.79$. (c) Dependence of $\mu_{h(4.3\ K)}$ on $\rho_{xx0(4.3\ K)}$ which shows $\mu_h$ is a decreasing function of $\rho_{xx0}$ at $T = 4.3$ K. (d) Charge carrier concentration, $n_h(T)$ for the same samples where (e) presents the enlarged view of $n_h(T)$ to depict minima ($T_{min}$) which occur nearly at the same temperatures as $T_{max}$ in $\mu_h(T)$ results. (f) $\mu_h(T)$ vs. $n_h(T)$ curves from $T \approx 4.3$ K to 300 K.

The analysis of possible scattering sources responsible for the mobility behavior is presented in Fig. 9(a) by fitting the data to inverse of the temperature. Since $\mu_h$ in such materials is suppressed by phonon scattering, the results could be fitted to $\mu_h \sim T^{-\gamma}$ where the exponent $\gamma$ is phonon related parameter and also it strongly depends upon the doping density due to the effect of dopants on the overall mobility results, see the increasing pattern of the obtained exponent with variation in impurity content presented in Fig. 9(c). The values of the parameter in order of increasing Sn/Mn impurities were estimated as, $\gamma \approx 0.46, 0.41, 0.43$ and $0.21$, respectively. Here, the $\gamma < 1$ range for GSMT alloys is significantly small compared to $\gamma = 2.1 - 2.6$ obtained for GaSe, MoSe$_2$ and MoS$_2$ and $\gamma = 1.2 - 1.7$ for MoSe$_2$ layers [76]. As discussed in detail by Fivaz et. al., several scattering sources could be responsible depending on the outcome of $\gamma$. The $\gamma = 1$ regime was assigned to the dominant role by acoustic modes whereas the variable values of $\gamma$ were ascribed to interaction with either homopolar or polar optical modes. In situation in which the center of mass of an atom and its mirror point remains unaffected and creates no first-order dipoles, such modes were categorized as homopolar whereas the opposite scenario produces polar modes. The mobility calculations by Fivaz et. al. showed that $\gamma$ shifted to lower than 1 regime for semiconductors like GaAs and InSb in the case of scattering of mobility carriers off the polar optical modes of the lattice [77]. In low dimensional systems like MoSe$_2$, the decline in $\gamma$ was attributed to the confinement of charge carriers in the layered structure [76]. However; the confinement of charge carriers could not be justified in the present bulk alloys which indicate that the small $\gamma$ values presumably originate from interactions with polar optical modes. In the low temperature regime where mobility drops faster from $T \approx 15 - 35$ K, the outcome of $\gamma$ still remains lower than 1 which means that the acoustic phonon scattering could be safely disregarded in the present results. According to previous explanations of carrier mobility in SCs materials, the $T^{-1.5}$ dependence is believed to indicate acoustic phonons as the dominant source [78] which indicates the acoustic phonon scattering can take values between γ = 1 [77] and γ = 1.5 [78]. While analyzing the current results, a very similar temperature dependence of $T^{-0.18}$ to $T^{-0.5}$ was reported by Irvine *et al.*, for lead halide perovskites [78] that could explain the origin of small exponent values. Although the exact origin is ambiguous, this unusual $T^{-0.5}$ dependence was proposed to arise due to the possible formation of large polarons [79,80]. As a final note, the significant drop in $\mu_h(T)$ and its negative temperature dependence indicates that the dominant phonon scattering takes over at higher temperatures [81]. Consequently, in view of previous similar interpretations, it is assumed that the dynamic disorder arising from higher level of optically excited phonons or presence of polarons could be the possible scattering sources [78–80]. The $\mu_h(T)$ results also show a large difference at $T \approx 4.3$ K for alloys of same material but unlike impurity contents. This is assumed to arise from residual resistivity due to substantial differences in the degree of static disorder [34]. Further, the reduction in $\mu_{h(4.3\ K)}$ shows a monotonic pattern as a function of $\rho_{xx0(4.3\ K)}$, see Fig. 9(c). For example the alloy with $\rho_{xx0(4.3\ K)}$ = 17 mΩcm delivers highest mobility of $\mu_{h(4.3\ K)} \approx 33$ cm$^2$/Vs which drops to ≈ 9 cm$^2$/Vs for $\rho_{xx0(4.3\ K)}$ = 72 mΩcm. The drop in $\mu_{h(4.3\ K)}$ is shown as a function of impurity content which reveals the mobility decreases for low impurity concentration. The reduced mobility could also derive from scattering at grain boundaries in polycrystalline materials.

Finally, variable temperature carrier concentration, $n_h(T)$ results are presented in Fig. 9(d,e). All four samples present high number of charge carriers of the order of $n_h \approx 1.5 - 3 \times 10^{21}$ cm$^{-3}$ at room temperature. The temperature dependence of $n_h(T)$ curves is reasonably strong below $T \approx 50$ K which then switches to a weakly dependent regime up to $T \approx 100$ K except for $x = 0.79$. The weak temperature dependence of $n_h(T)$ at $T \geq 100$ K demonstrates degenerate semiconductors like behavior [23]. At $T \approx 8.8$ K and ≈13.1 K, minima (shown as $T_{min}$) were recorded which are closely related to the $T_{max}$ values in mobility curves, Fig. 9(e). Also, the $n_h(T)$ results vary as a function of impurity content and decrease for higher Sn+Mn ions in the GeTe host lattice particularly at high temperatures. The mobility results are next shown against carrier concentration in an attempt to understand a correlation between both quantities; refer to Fig. 9(f). The data manifest that $\mu_h(n)$ is continuously decreasing except for the sample with $x = 0.79$ where a hump occurs at $T \approx 30$ K. This clearly illustrates that the charge scattering in the high temperature and carrier concentration regime significantly influence the mobility values. Another visible feature is that the mobility seems to exhibit less dependence on carrier concentration at certain point e.g. between $n_h(T) \approx 1.4 - 3 \times 10^{21}$ cm$^{-3}$. The decline in $\mu_h(n)$ for higher $T$ is assumed to develop due to ionized sites which act as scattering centers in the lattice at elevated temperatures. In such a scenario, the presence of Ge vacancies and ionized impurities considerably decrease the carrier mobility due to Coulomb scattering centers which impacts the carriers' trajectories [82].

**Conclusions**

The Mn doped Ge$_{1-x}$Sn$_x$Te multiferroics were studied which showed metallic-like behavior in the temperature dependent resistivity results. From temperature dependent transport data, the true nature of scattering processes cannot be claimed due to the unusual fitting outcomes. The analysis of the $\rho_{xx}(T)$ results reveal diverse scattering mechanisms responsible ranging from static to dynamic disorder however; in the high temperature regime, the unusual scaling results could not be interpreted in terms of only phonons. The assumed understanding of the carrier mobility on the other hand finds traces of polarons at high temperature.

Similarly, extrinsic anomalous Hall effect is obtained from the high field magnetotransport data. The AHE of the GSMT alloys falls into the low conductivity hopping regime which excludes the possibility of intrinsic mechanism. After the scaling analysis with various relations, the possibility of a single dominant scattering source is disregarded since the results cannot be scaled either with only skew or side jump relations. In order to examine the approximate contribution of each scattering mechanism to AHE, the interpretation of anomalous Hall resistivity identifies its scattering origin in the superposition of extrinsic sources. Although the scaling results hint towards superposition of skew scattering and side jump, the tendency of the obtained parameters signifies that a large share of contribution comes from the later source. The dominancy of side jump in the GSMT samples could also be anticipated due to comparable mean free path of charge carriers to the inter Mn distances in IV-VI DMSs. Finally, the negative magnetoresistance interpretation demonstrates a spin-disordered state however; the interaction among spins in such a state below $T_C$ is different from pure spin-glass systems.

ACKNOWLEDGMENTS

The research was financed by the National Science Centre, Poland under the project number 2018/30/E/ST3/00309.
Corresponding author: akhaliq@ifpan.edu.pl